# Reflection compensation mediated by electric and magnetic resonances of all-dielectric metasurfaces

Viktoriia E. Babicheva,[1,2,*] Mihail I. Petrov,[1,3] Kseniia V. Baryshnikova,[1] and Pavel A. Belov[1]

[1]ITMO University, Kronverkskiy, 49, St. Petersburg 197101, Russia
[2]Center for Nano-Optics, Georgia State University, P.O. Box 3965, Atlanta, GA 30302
[3]Department of Physics and Mathematics, University of Eastern Finland, Tulliportinkatu 1, Joensuu 80101, Finland
* Corresponding author: v.babicheva@phoi.ifmo.ru

**All-dielectric nanostructures have recently emerged as a promising alternative to plasmonic devices, as they also possess pronounced electric and magnetic resonances and allow effective light manipulation. In this work, we study optical properties of a composite structure that consists of a silicon nanoparticle array (metasurface) and high-index substrate aiming at clarifying the role of substrate on reflective properties of the nanoparticles. We develop a simple semi-analytical model that describes interference of separate contributions from nanoparticle array and the bare substrate to the total reflection. Applying this model, we show that matching the magnitudes and setting the π-phase difference of the electric and magnetic dipole moments induced in nanoparticles, one can obtain a suppression of reflection from the substrate coated with metasurface. We perform numerical simulations of sphere and disk nanoparticle arrays for different permittivities of the substrate. We find full agreement with the semi-analytical results, which means that the uncoupled-element model adequately describes nanostructure reflective properties, despite the effects of induced bi-anisotropy. The model explains the features of the reflectance spectrum, such as a number of dips and their spectral positions, and show why it may not coincide with the spectral positions of Mie resonances of the single nanoparticles forming the system. We also address practical aspects of the antireflective device engineering: we show that the uncoupled-element model is applicable to the structures on top of silicon substrates, including lithographically defined nanopillars. The reflectance suppression from nanoparticle array on top of the silicon substrate can be achieved in a broad spectral range with disordered nanoparticle array and for a wide range of incidence angles.**

## 1. Introduction

A variety of plasmonic nanostructured coatings were proposed to manipulate light at the nanoscale (see reviews[1-3] and references there), with particular interest to control reflection from the material interfaces[4,5] and improving photovoltaic properties.[6,7] High-refractive-index nanoparticles are considered as one of the most promising alternatives to plasmonic nanostructures to be applied in photovoltaic devices and ultra-thin functional elements, so-called metasurfaces.[8-10] With the typical linear dimensions of 50-200 nm, such nanoparticles allow Mie resonances at visible spectral range and possess induced electric and magnetic moments.[11-15] It gives an opportunity to obtain magnetic optical response without using metal inclusions such as split-ring resonators and avoid additional Ohmic losses associated with metal. Such all-dielectric nanostructures, for instance nanoparticle arrays, demonstrate a variety of unique effects, including suppression of reflection in particular directions,[16,17] generalized Brewster effect,[18] Raman scattering enhanced by magnetic resonances,[19] bi-anisotropy,[20] and substrate-induced bi-anisotropy,[21,22] which can be useful for efficient light harvesting[23] or to improve performance of ultra-thin elements based on metasurfaces.[24]

To suppress scattering in a backward direction from a single nanoparticle in air, the first Kerker condition should be satisfied,[25] that is electric and magnetic moments should be equal in magnitude and phase. As a result, for spherical nanoparticle array, antireflective properties are observed at wavelength either larger than the wavelength of the magnetic resonance or smaller than the wavelength of the electric resonance.[16,26] Designing nanoparticle geometry, one can obtain spectral overlap of electric and magnetic dipole resonances.[27] There is a number of studies that suggest using nanoparticles with more complex shape, such as disks[26,28-31] or cubes[32], or moderate refractive index[33] to decrease backscattering in a broad spectral range. Alternatively, high reflection between the resonances can be utilized for developing a perfect reflector based on all-dielectric metasurface (for near-infrared range designs see e.g. refs. [34,35]). Most of these studies were performed with nanoparticles in air[32] or metasurfaces placed on top of the low-index substrate and embedded with the material of similar refractive index[29] to avoid the influence of substrate. However, in many cases, nanoparticles should be placed on top of other material (experiments with the substrates[36] or designing metasurfaces[37,38] and photovoltaic elements[23,39-40]), and one needs to understand optical properties of these structures.

The change of scattering properties of nanoparticles on top of the substrate, including directionality, was studied in detail earlier[26,41,42] and received more attention for both plasmonic[43] and dielectric nanoparticles[44,45] very recently. In contrast, only a few works[46-49] experimentally investigate the problem of nanoparticle arrays on the substrate, but none of the works addresses the important role of interference of nanoparticle multipole scattering waves with the wave reflected from the substrate. In particular, the experimental results reported in Ref. [48] show the reflectance suppression from silicon/germanium nanoparticles over silicon/germanium substrate but disregard the role of the substrate. In the present work, we show that there can be one or two dips in reflectance, and their positions are varied with the period on the array and do not necessarily coincide with the wavelength of Mie resonances of single nanosphere.

We propose a semi-analytical model separating contributions from the nanoparticle array and the bare substrate and compare it to whole-structure numerical simulations obtaining good agreement. It allows demonstrating that reflection suppression occurs because of the destructive interference between waves scattered by electric and magnetic dipole moments and the wave reflected from the substrate itself. Even though backscattering from the nanoparticle array (that is reflection) is high, one can obtain a near-zero total reflection from the structure by destructive matching the metasurface-reflected wave with the substrate-reflected wave. We emphasize that the good agreement between whole-structure numerical simulations and semi-analytical results means that the simple uncoupled-element model is sufficient to describe reflection spectra of the silicon particles on high-index substrate and, in particular, substrate induced bi-anisotropy[44] does not play any significant role.

We perform calculations for different permittivities of the substrate and specify what properties the structure should possess to suppress reflection from the highly reflective substrate. Specifically, we show that for nanoparticle array on the substrate with permittivity $\varepsilon_s \gtrsim 4$, strong suppression of reflection occurs spectrally between the magnetic and electric dipole resonances, which is reverse to the case with homogeneous embedding, where the wavelength of near-zero reflectance is always from one side of both resonances.[16,26] We also show good agreement between numerical and semi-analytical models for both sphere and disk nanoparticle arrays, although, in the case of disks, the coupling of resonant mode to the substrate is strong and requires an additional low-index layer. The observed anti-reflectance effect is a cumulative action of array elements – each nanoparticle functions as anti-reflective unit – rather than collective grating effect. This property allows achieving broadband anti-reflectance of a silicon substrate with disordered nanosphere array.

The paper is organized as follows. In Sec. 2, we propose a semi-analytical model for nanoparticle array over the dielectric substrate with arbitrary refractive index, which treats metasurface and substrate as optically uncoupled elements. With this model, we explain the observed reflectance suppression and confirm it by the numerical simulations. In Sec. 3, we study the antireflectance effect of silicon metasurface on top of the silicon substrate to show the potential applicability of the system for photovoltaics and optoelectronics. In Sec. 4, we show that the observed antireflectance effects are tolerant to disorder in the system. In Sec. 5, we discuss all-dielectric metasurface based on silicon nanopillars, and we demonstrate the possibility of decoupling by low-index buffer layer beneath the metasurface. In contrast to spherical particles such structures can be fabricated by common planar methods and, thus, are of special interest. In Sec. 6, we discuss the applicability of the model and manifestation of its results in recent experiments. In Sec. 7, we summarize results and draw conclusions.

## 2. Nanoparticle array on high-index substrate: semi-analytical model

To start with, we consider silicon nanosphere array (permittivity data is from Ref. [50]) on the substrate with permittivity $\varepsilon_s$ varied between 1 and 20 [Fig. 1(a)]. We perform numerical simulations (CST Microwave Studio) of an array consisting of the nanospheres with radius $R = 60$ nm arranged in a simple square lattice with a period of $d = 335$ nm. Nanoparticles are placed right on top of the substrate, there is neither superstrate nor top coating, and none of the nanoparticle's part is embedded. The chosen parameters allow observing of anti-reflectance effects in the region of silicon absorption, which can find applications in photovoltaics and optoelectronics. However, the effects can be also observed in other spectral ranges due to the scalability of Mie resonance wavelength with nanoparticle size. In this paper, we mainly consider the normal incidence of the light (figures where the angle of incidence is not mentioned), but we also show that the studied effect can be observed at oblique incidence. Throughout the text, we use the term 'metasurface' for nanoparticle array without the substrate and 'whole structure' for nanoparticle array on top of the substrate.

Absorbance spectra show two distinct peaks on the wavelengths $\lambda_{ED} = 435$ nm and $\lambda_{MD} = 517$ nm, which correspond to electric (EDR) and magnetic dipole resonances (MDR), respectively [Fig. 1(c)]. Both resonances result in dips in transmittance spectra [Fig. 1(b)]. An increase of $\varepsilon_s$ causes a slight non-monotonic changes in both peaks: they have a minimum height when $\varepsilon_s = 4$-6 followed by an increase up to $\varepsilon_s = 20$. Slight variations of absorption peaks' position and intensity occur because polarizability of each nanoparticle is affected by the presence of substrate and by the scattered fields of other particles summed over the lattice. Alternatively, the variations can be explained as a result of bi-anisotropy effect[44]. Still, the weak coupling of sphere modes to the substrate[26] makes the insignificant influence of the substrate on the nanoparticles' absorbance resonances. For the same reason, we do not expect strong lattice resonances or diffraction effects on $\lambda_{eff} = \lambda/\sqrt{\varepsilon_s} \approx d$ as the field scattered inside the substrate is negligibly low. For $\lambda < 380$ nm and $\varepsilon_s = 1$, one can observe additional peaks, which are also EDRs and occur because of the high dispersive silicon permittivity, but we do not consider these peaks in this work because they are out of the visible spectral range.

In contrast to absorbance, reflectance profile varies strongly [Fig. 1(d)]. For $\varepsilon_s = 1$ we see two peaks, which correspond to EDR and MDR, the slight decrease between EDR and MDR, and two dips at $\lambda_{K1} = 559$ nm and $\lambda_{K2} = 417$ nm with near-zero reflectance. The dip at $\lambda_{K1}$ disappears already for small substrate permittivity, e.g. $\varepsilon_s = 2.5$. The most striking feature is a decrease of reflectance between EDR and MDR down to a value close to zero for $\varepsilon_s \approx 4$-6. It is the most pronounced for $\varepsilon_s = 4$ (denoted as $\varepsilon_{s1}$) and observed at the wavelength $\lambda_A = 482$ nm. Upon further increase of $\varepsilon_s$, one can see the splitting of the dip into two dips spectrally close to EDR and MDR.

Let us explain the phenomenon of the change of the antireflection point with substrate permittivity increase in detail. Without the substrate (i.e. $\varepsilon_s = 1$), at $\lambda_{K1}$ and $\lambda_{K2}$ the first Kerker condition is satisfied,[25] that is $\mathbf{p}$ and $\mathbf{m}$, electric (ED) and magnetic dipole (MD) moments induced in the nanoparticles, are in phase (zero phase difference) and equal in magnitude, which suppresses backscattering and allows high transmission. At $\lambda_{K1}$ and $\lambda_{K2}$ the dipole moments are in-phase, Poynting vector of the field scattered by dipoles $\mathbf{S} \propto \left[ \mathbf{p} \times \mathbf{m} \right]$ points down, which means that the dipole efficiently scatters waves in the forward direction, and no reflection is observed (see K1 and K2 points in Fig. 2 and Fig. S2(a)-(d) in Ref. [51]). Furthermore, we defined anti-Kerker (AK) point as the wavelength of the highest forward-to-backward ratio. For the wavelength between EDR and MDR, $\mathbf{p}$ and $\mathbf{m}$ have π phase difference, Poynting vector $\mathbf{S}$ points up (see AK point in Fig. 2 and Fig. S2(e),(f) in Ref. [51]), and reflection from the metasurface is increased [$\lambda \approx 440$-510 nm, $\varepsilon_s = 1$ in Fig. 1(d)].

In contrast to the homogeneous environment ($\varepsilon_s = 1$), in case of the high-index substrate, there is a strong reflection from the substrate surface. From the absorbance spectra [Fig. 1(c)], we know that substrate presence weakly affects ED and MD induced moments, and one can expect that approximately the same phase difference between $\mathbf{p}$ and $\mathbf{m}$ oscillations will be in the case of $\varepsilon_s \neq 1$. Indeed, simulations show that for $\varepsilon_s = \varepsilon_{s1}$ dipole moments have a π-phase difference at $\lambda_A = 482$ nm, and correspondingly their Poynting vector points up (Fig. S2(g),(h) in Ref. [51]), and reflection from the nanoparticles is increased. This motivates us to consider a semi-analytical model, where the metasurface and substrate are considered independently (often referred to as *decoupled system*[52]).

The optical response of the metasurface consisting of interacting electric and magnetic dipoles can be described with reflection and transmission Fresnel coefficients (see more details in refs. [11,29]):

$$r_{MS} = \frac{ik_0}{2d^2}\left(\alpha_e^{eff} - \alpha_m^{eff}\right), \quad t_{MS} = 1 + \frac{ik_0}{2d^2}\left(\alpha_e^{eff} + \alpha_m^{eff}\right), \quad \textbf{(1a)}$$

where $k_0 = 2\pi/\lambda$ is the free-space wavenumber, $\alpha_e^{eff}$ and $\alpha_m^{eff}$ are the effective electric and magnetic polarizabilities that take into account interaction between the nanoparticles in the lattice, and $d$ is the array period. The electric field reflected from the metasurfaces is $E_{MS}^r = r_{MS} E_0$, where $E_0$ is the field amplitude of normally incident plane wave calculated at the plane $z = 0$ ($E_0$ is complex with absolute value equal to field intensity and complex argument denoting the phase). Thus,

$$E_{MS}^r = E_{MS,e}^r + E_{MS,m}^r, \quad \textbf{(1b)}$$

where

$$E_{MS,e}^r = \frac{ik_0}{2d^2}\alpha_e^{eff} E_0 \text{ and } E_{MS,m}^r = -\frac{ik_0}{2d^2}\alpha_m^{eff} E_0 \quad \textbf{(1c)}$$

are the electric field components caused by electric and magnetic dipole scattering, respectively.

The reflectivity of the dielectric substrate is characterized by its own Fresnel coefficient, which we denote as $r_s$. Assuming that the metasurfaces and the substrate act as independent in-line optical elements, we can calculate the field reflected from the substrate at the plane $z = 0$ [Fig. 3(a)] as Fabri-Perot type series:

$$E_s^r = t_{MS}^2 r_s e^{2ik_0 R} E_0 + t_{MS}^2 r_{MS} r_s^2 e^{4ik_0 R} E_0 + \ldots = \\ = t_{MS}^2 r_s e^{2ik_0 R} / (1 - r_{MS} r_s e^{2ik_0 R}) E_0. \quad \textbf{(2)}$$

Finally, the total reflected field can be calculated by summing up the fields reflected directly from the metasurfaces and from the substrate:

$$E_{tot}^r = E_{MS}^r + E_s^r. \quad \textbf{(3)}$$

Using formula (3) for reflected field amplitude, one can express reflectance simply as $|E_{tot}^r / E_0|^2$. To calculate the field $E_{tot}^r$, we extract the metasurface Fresnel coefficients $r_{MS}$ and $t_{MS}$ from the numerical simulations of metasurfaces without the substrate and $r_s$ from simulations of bare substrate separately. Alternatively, one can find $r_s$ as $r_s = (1-\sqrt{\varepsilon_s})/(1+\sqrt{\varepsilon_s})$. From Fig. 1(d), one can see good agreement between the results of whole-structure simulations and calculations with the model (3), and in particular, the existence of anti-reflectance point at $\lambda_A \approx 482$ nm for the optimum value of $\varepsilon_s = \varepsilon_{s1}$. Further, increasing the substrate permittivity, we observe a split of the anti-reflectance point into two points, which are spectrally close to the EDR and MDR and also present in Fig. 1(d) for the whole-structure simulations of nanoparticle array with the substrate.

On a qualitative level, the antireflective effect observed between the EDR and MDR can be understood from the diagram shown in Fig. 3(b),(c). The phase of polarizabilities $\alpha_e^{eff}$ ($\alpha_m^{eff}$) gradually changes from 0 to π as the wavelength is passing EDR (MDR) from long-wavelength to short-wavelength region. Thus, between the EDR and MDR, $\alpha_e^{eff}$ is almost in-phase with the incident wave $E_0$, and $\alpha_m^{eff}$ has almost π-phase relative to $E_0$. According to Eq. (1c), the ED-backscattered field $E_{MS,e}^r$ has a π/2 phase shift relative to the incident wave $E_0$ [shown with a purple arrow in Fig. 3(b)]. Similarly, according to Eqs. (1b,c), the electric field $E_{MS,m}^r$ is shifted less than π/2 relative to the incident wave $E_0$ [shown with a green arrow in Fig. 3(b)]. We note that our numerical simulations do not allow to separate ED and MD contributions in the total field at $\lambda_A$, and consequently, exact complex values of $E_{MS,e}^r$ and $E_{MS,m}^r$ components are unknown and drawn schematically.

The wave reflected from the substrate $E_s^r$ is shown by the blue arrow in Fig. 3(b),(c). For the sake of simplicity, in the diagram, we plot only the first term in series of Eq. (2). One can show that single-reflection approximation gives good agreement with the whole-structure simulations (Fig. S3 in Ref. [51]). In this case, $E_s^r$ has π phase shift relative to the incident field due to the reflection from optically dense substrate shown by black dashed arrow. Additional phase incursion of $2k_0 R$ is acquired due to light passing from the plane of nanoparticles' center to the substrate surface and back [see side view in Fig. 3(a)], and this phase accumulation approximately equals to π/2 for the considered parameter range (it is exactly π/2 for $R = 60$ nm and $\lambda = 480$ nm). Finally, because EDR and MDR are spectrally well separated and nanoparticle array is sparse, the component $t_{MS}^2 \approx 1$ between the EDR and MDR [Fig. 1(b)]. This gives the total approximately π phase shift between the $E_{MS}^r$ and $E_s^r$, which leads to destructive interference of waves reflected from the metasurface and the substrate.

Overall, one can see that high backscattering from metasurface enables near-zero reflection from the whole structure (nanoparticle array on the substrate) once the metasurface-reflected wave is matched by the substrate-reflected wave. Increasing the substrate index (*e.g.* $\varepsilon_s = 20$), we obtain the higher amplitude of the reflected wave according to Fresnel formula [bigger inner circle in Fig. 3(c)], and complete compensation of the substrate-reflected wave at the same wavelength by the metasurface-scattered wave is not possible. The minimum reflection is shifted toward EDR and MDR ($\lambda_{A1} = 438$ nm and $\lambda_{A2} = 504$ nm at the curve corresponding to $\varepsilon_s = 20 \equiv \varepsilon_{s2}$). At these points, closer to resonances, the amplitude of the wave backscattered from the metasurface is larger and the compensation is better.

The period of the nanoparticle array drastically influences the amplitude of metasurface-reflected wave according to Eq. (1a). Under the study, we carried out numerical simulations for different periods of the array: $d = 335$ nm is chosen as the most representative case, but other periods were also studied [51]. To suppress reflection from low-index substrate with $\varepsilon_s \simeq 2$-3, one requires a sparse structure, *e.g.* $d = 417$ nm (Fig. S5 in Ref. [51]); for higher permittivity of substrate $\varepsilon_s \simeq 4$-6, array period $d = 335$ nm is the most suitable (see Fig. 1); and for higher permittivity of substrate $\varepsilon_s \simeq 7$-20, dense metasurfaces are required with period $d \simeq 250$ nm (Fig. S4 in Ref. [51]). Thus, the structure possesses a high tunability, and appropriate period always can be found to match the substrate permittivity.

In the recent work,[44] silicon spherical particle with the similar size ($R = 65$ nm) on top of different substrates was studied, and the role of substrate-induced bi-anisotropy in the spectral shifting of resonances was analyzed. It was shown that silicon substrate modifies extinction cross-section of the nanosphere, and in particular, the extinction cross-section becomes two times higher at EDR. This change of extinction cross-section may explain deviations between our semi-analytical model and numerical calculations, although does not significantly affect the reflectance properties.

## 3. Nanosphere array on the silicon substrate

To address practical aspects of the anti-reflectance effect under consideration, we perform calculations of nanosphere array ($R$ = 60 nm) on the silicon substrate. In this analysis, the permittivity of silicon substrate is the same as for nanoparticles (data from Ref. [50]). Here, we study nanoparticle arrays of different periods and show that the anti-reflectance can be observed in a wide range of array periods. Moreover, comparison of numerical simulation results and calculations according to semi-analytical model shows very good agreement [Fig. 4(a),(b)], including periods down to 130 nm (note that nanosphere diameter is $2R$ = 120 nm). One can distinguish the presence of three different regimes: (i) for $d$ = 130-180 nm, there are one dip spectrally close to EDR and one broad peak between EDR and MDR; the dip is explained by the destructive interference of ED with substrate-reflected wave; (ii) for $d$ = 180-250 nm, there is one broad dip between EDR and MDR; and (iii) for $d$ = 250-400 nm, there are two dips in reflectance. One can also see a diffraction resonance at $\lambda_d = d$, and its peaks and dips are not quantitatively described by the semi-analytical model. The destructive interference of different multipoles with a substrate-reflected wave can be seen for almost all periods as a narrow dip at a wavelength approximately 360 nm.

Due to the interference nature, the anti-reflectance effect can be observed for nanoparticle array on Si substrate for different angles of light incidence ($R$ = 60 nm, $d$ = 335 nm, Figs. 5 and S6 in Ref. [51]). Both TE (electric field is parallel to the substrate) and TM polarizations (the magnetic field is parallel to the substrate) are studied. In TE polarization (Fig. S6(a),(b) in Ref. [51]), the spectrum profiles are similar in a broad range of angles, i.e. for $\alpha < 60°$, which is expected for the wave with in-plane electric field. In TM polarization [Fig. 5], the reflectance from the substrate itself decreases with the increase of the angle, thus the spectrum changes from the two-dip profile ($\alpha < 30°$) to one-dip profile ($30° < \alpha < 60°$), which is similar to the effect observed for moderate-refractive-index substrates in Fig. 1(d) or for small-density arrays ($d$ = 130-180 nm) in Fig. 4(a). This polarization is also more affected by diffraction effects: note profile modification in the proximity of red solid line in Fig. 5 and compare with Fig. S6(a) in Ref. [51].

## 4. Disordered nanoparticle array

From the application point of view, the disordered structures can be more beneficial than periodic ones. The self-assembled and disordered structures have been already suggested for possible application in photovoltaics,[53,54] which stimulated us to consider disordered metasurface here. We also note that fabrication of such structures does not require lithography process and thus can be less expensive in comparison to the fabrication of ordered nanoparticle array. In particular, there are a few methods[55-57] of fabricating structures that do not require precise but expensive optical or electron beam lithography and post-processing. As an alternative to lithography, the cost-effective self-assembly methods have attracted an interest of researchers and their industrial partners. The disadvantage of self-assembly is the certain amount of disorder introduced into the fabricated structures, which is related to the stochastic character of the process. The known self-assembly methods of fabricating of large-area spherical silicon nanoparticles arrays are based on colloidal[58] and dewetting[59,60] techniques. Moreover, recent advances of spin-coating methods[61,62] can be used for simple fabrication of disordered nanoparticle arrays utilizing commercially available nanoparticles[63,64].

The anti-reflectance effect observed in wide angle range is determined by the directivity diagram of each individual nanoparticle in the array. In general, nanoparticle resonance positions are affected by the presence of substrate and neighboring particles in the array, but for the structure under consideration, both effects are weak. Small shifts of EDR and MDR spectral positions with the change of substrate permittivity [e.g. Fig. 1(c)] or inter-particle distance (e.g. Fig. 4) indicate their weak influence. Thus, in the first order approximation, the array can be considered as a set of non-interacting nanoantennas, which locally suppress the substrate reflection (experimentally observed in Ref. [48]). Consequently, one can expect that the observed anti-reflectance effect will persist in the absence of order and periodicity and that the general optical properties of disordered nanoparticle array will remain similar to the one with periodicity.

We numerically calculated reflectance from the silicon surface coated by 49 disordered nanospheres (Fig. 6) with radiuses $R$ uniformly distributed in 50 to 80 nm range $R$ = 50-80 nm (see inset in Fig. 6) on the area 1.7 x 1.7 $\mu m^2$ (approximately 240 x 240 $nm^2$ per nanoparticle). MDR and EDR of nanospheres with such radiuses are excited in a broad spectral range (for scattering and absorption cross-sections, see Fig. S7 in Ref. [51]). Because of the high reflection from each nanosphere and its destructive interference with the substrate-reflected wave, the total reflection from the structure is significantly reduced. In the spectral range 300-800 nm, reflection from silicon surface with the nanosphere coating is less than from silicon surface with a 55-nm-thick $Si_3N_4$ layer, which is optimal for zero-reflection at 470 nm (single quarter-wavelength coating). Overall, the reflectance is reduced down to 10% at wavelengths 300-650 nm.

We did not optimize structure parameters, but according to performed calculations (not shown here), further decrease of reflectance at broad dip (wavelengths 380-650 nm) can be effectively controlled by varying nanoparticle sizes and density. Because of the spherical shape of nanoparticles, the anti-reflective effect most likely remains the same under the oblique incidence of light in a broad range of angles (similar to what we showed for the periodic array, Figs. 5 and S6 in Ref. [51] and Ref. [65]).

One can mention that smearing out of resonances in a similar nanostructure (silicon disk array[24]) caused an increase of transmission through the array because of the better overlap of EDR and MDR. Thus, the effect of disorder can be counterintuitive, and the case of the present paper, that is metasurface in the non-homogenous environment, requires more study on this matter. Overall, size and the average density of nanoparticles are parameters that influence the most, while the precise position of nanoparticles affect less. Slight disorder of nanoparticles in terms of size and distance between them may cause smearing out of resonances, but our approach still can be applied using metasurface reflection and transmission coefficients with account of disorder.[66,67]

## 5. Nanopillar array

Today, lithography is one of the most common methods of fabrication of planar structures and metasurfaces, and the vast majority of practical applications of all-dielectric metasurfaces has been proposed based on lithographically defined nanoparticle arrays. In particular, the typical meta-atom is silicon nanopillar on top of the low-index substrate or in the homogeneous low-index matrix. Placing silicon nanopillars directly on top of high-index substrate suppresses their resonances due to mode leakage into the substrate[26]. In what follows, we propose the structure where the metasurface properties are preserved despite the lager contact area of the nanoparticle with the high-index substrate. To avoid leakage of nanoparticle modes, one needs to separate it from the high-index substrate by a thin low-index intermediate layer.

We study the structure that consists of a periodic array of nanodisks on top of the SiO$_2$ buffer layer of varied thickness $s$ [Fig. 7(a)]. The radius of the nanodisks is $R = 50$ nm, height is $h = 120$ nm, and the period of the structure is $d = 350$ nm. In contrast to the previous works where nanodisk dimensions were chosen to maximize the overlap of the resonances,[24,28,29] our dimensions enable distinct ED and MD resonances, i.e. $\lambda_{ED} = 440$ nm and $\lambda_{MD} = 510$ nm [Fig. 7(b)]. By including of buffer layer, we also suppress wave propagation inside the substrate and eliminate diffraction effects.[68]

In the previous sub-section, we developed a semi-analytical model for the nanoparticle array on infinitely thick high-index layer (substrate). Optical properties of nanoparticle arrays can be studied by different techniques (see e.g. ref. [69] and recent review ref. [70]), and our semi-analytical model can be directly deduced from the standard transfer-matrix method where light propagation through each element is considered separately (decoupled system[52]). The transfer-matrix method can be used to take into account additional reflections from the buffer layer and obtain the Eq. (3) for multilayer structure {metasurface / air gap / SiO$_2$ buffer layer / Si substrate} [51].

For the $s > 15$ nm, both MDR and EDR resonances are well defined [Fig. 6(c)-(g)] and agreement between the numerical simulations and semi-analytical calculations of reflectance spectrum are good [Fig. 6(h),(i)]. Although the reflectance dip associated with MD resonances disappears for $s > 30$ nm, it is well described by the decoupled-system model and related to interference of waves reflected from buffer layer.

To sum up, we prove that one can achieve a zero reflectance from the highly-reflective substrate with nanoparticle coating for the wavelength between EDR and MDR of nanoparticles. The effect is observed because of the destructive interference of waves reflected from the metasurface and from the substrate, which is confirmed by the semi-analytical model. We numerically demonstrate a possibility of broadband anti-reflectance effect (300-800 nm wavelength) with a disordered-nanoparticle coating of the silicon substrate. Finally, we show that for nanodisk array on the low-index buffer layer of 15-30 nm thickness, modes do not leak out to the silicon substrate and decoupled-system model reasonably good describes antireflective properties of the metasurface.

## 6. Discussion

### A. Applicability of the semi-analytical model

The proposed model of separated contributions shows surprisingly good agreement with the whole-structure simulation results for nanosphere metasurface. In this case, a slight discrepancy between the results of whole-structure simulations and model calculations is observed at the wavelength close to $\lambda_{ED}$ and $\lambda_{MD}$: the model gives increased reflectance in comparison to whole-structure simulations profiles. This discrepancy may come from the substrate-induced bi-anisotropy[44], which gives significant contribution at the wavelength of EDR and MDR. However, the agreement of the results of whole-structure simulations and model calculations also confirms that the role of the substrate induced bi-anisotropy for spherical dielectric nanoparticles is weaker than for plasmonic nanoparticles on the high-index substrate[22,71,72]. Similar calculations for silver nanoparticle array on silicon substrate do not give good agreement with the numerical calculations (not shown here), which means that the interaction with the substrate is very strong. In contrast, results of this work show that all-dielectric metasurfaces can be treated as an uncoupled from the substrate element, thus, allowing to combine them with other optical elements preserving their optical properties.

One can draw a parallel between the studied problem of antireflection and designing perfect absorbers[5,52,73,74] as in both cases the reflection needs to be suppressed, but for the perfect absorbers, transmission through the structure needs to be zero as well. Similar decoupled-system model was used in ref. [52] resulting in a conclusion that interference between the reflected wave and electric resonances can suppress the total reflection from the metasurface.

Recently, a generalized problem of reflection suppression by resonant nanoparticles on the substrate was considered in ref. [69] taking into account multiple diffraction orders. While we study subwavelength periodic array and calculate only zero diffraction order in visible spectral range, in contrast to ref. [69], we analyze a structure with two dipole resonances and show that interplay between their phases is important for obtaining destructive interference and antireflection.

For metasurfaces in vacuum, the equality of ED and MD moments means that effective impedance of the metasurface – proportional to $\sqrt{(\mu/\varepsilon)}$ – is equal to the vacuum impedance, see e.g. ref. [28]. This ensures transmission through the structure in vacuum without reflection, which was studied in the original work by Kerker et al.[25] The studied antireflection effect for the nanoparticle array over the substrate shows the necessity of impedance matching between the metasurfaces and the high-index substrate. It can be referred to as *anti-phase Kerker effect*, being a counterpart of Kerker effect for the metasurface in vacuum, where the anti-reflectance is observed for in-phase resonances condition.

### B. Structure properties

As was predicted recently,[75] one can observe a highly-directional scattering from single nanoparticle provided that higher-order moments are balanced (e.g. ED and electric quadrupole in the case of plasmonic nanoantenna[75]). The generalized approach to obtain zero backscattering/reflectance is (i) to operate in high-scattering regime (proximity to EDR and/or MDR) and (ii) to obtain destructive interference of this scattered field with the wave reflected or scattered by another element in the structure.

It is important to stress that the studied anti-phase Kerker effect can be observed only for nanostructures with the following properties:

(A) *Shape*. Coupling between the nanoparticles and the substrate should be weak. Otherwise, nanoparticle modes leak out to the substrate upon increase of its refractive index and uncoupled approach fails. We show good agreement of the model for silicon nanospheres, but it works poorly for nanopillars without silica buffer layer Ref. [51].

(B) *Size*. Dimensions of the nanoparticle along light-incidence direction should approximately satisfy the equation $a = \lambda_A/4$, where $\lambda_A$ is the wavelength of near-zero reflectance (for instance, we studied spheres with diameter $a = 120$ nm and observed near-zero reflectance at $\lambda_A = 482$ nm). If this condition is not satisfied, the substrate-reflected wave does not accumulate π phase shift relative to the wave scattered by ED and MD and destructive interference is not achieved (see Fig. 3(b) for a detailed explanation).

(C) *Density*. For a substrate with given refractive index $\varepsilon_s$ one can tune the filling factor of metasurface to meet the anti-reflectance condition. From one side, for the high-index substrate, the density of nanoparticles in the metasurface array should be high to compensate the strong reflection from the substrate. From another side, nanoparticle array should be sparse enough so that transmittance through metasurface $T \gtrsim 0.5$ (for considered nanostructures $T \approx 0.65$-$0.8$). Here, we should note, that there are several limitations on tuning the metasurface filling factor: (i) Eq. (1a) are not valid for very dense arrays when the dipole model breaks down;[11] In particular, because of the higher wavelength and limitations described in ref. [11], we do not observe MDR resonances for $d < 230$ nm, which agrees with the analytical calculations for the similar structure[11] (ii) MDR appears to be much more sensitive to the density of nanoparticles [Fig. 4(c)] and angle of light incidence (Fig. S6 in Ref. [51]) than EDR; (iii) compensating the reflectance from low-index substrate one needs to use sparse arrays with large period, which can be comparable to wavelength, breaking the non-diffraction condition in air; and (iv) in contrast, even for the case of 20-30 nm thick high-index layer on top of the low-index bulk substrate, dense metasurface is required to compensate substrate-reflected wave.

We emphasize that the key factor for agreement of numerical simulations and semi-analytical model results is a separation of MDR and EDR [Fig. 4(c)] rather than low density of nanoparticles. From Fig. 4(d), one can see that transmittance is low for $d < 200$ nm, and nevertheless, the agreement between simulations and model is good.

(D) *Scalability.* In general, the effect is scalable and can be observed at another wavelength range provided that the structure is properly designed, i.e. conditions (A)-(C) are satisfied. In particular, Mie resonance in the dielectric sphere is excited at $\lambda_{ED} \approx n_p a$ ($n_p$ is the refractive index of the particle), which is spectrally close to the wavelength $\lambda_A = 4a$ [defined in (B)] and thus both effects occur together. Moreover, near-zero losses of silicon in near-infrared range make this material more favorable in comparison to plasmonic structures.

In contrast to the present work, where we aim to improve the antireflective properties of metasurfaces, one can utilize nanoparticle arrays for perfect reflectors. This can be done by using very dense arrays (see (C) condition) or changing the height of the nanostructure (see (B) condition). One of the demonstrations of the perfect reflector in near-infrared spectral range based on all-dielectric metasurface was reported in ref. [34], where particle array is dense, condition (C) is not fulfilled, and therefore the whole structure possesses high reflection.

### C. Reflectance spectral profile

Some confusion in the literature[46-49] appears in the interpretation of experimental results with reflection dips for dielectric nanoparticles on high-index substrates because of the proximity to Mie resonances and/or absence of analytical solutions of the problem. In the present work, we clearly show that anti-reflection effect appears in the proximity of multipole resonances of silicon nanoparticle, but cannot be ascribed to the resonances themselves. In Fig. 4(a), we show that reflectance spectra have one or two dips, and their positions are drastically different for different periods on the array and do not necessarily coincide with the wavelength of Mie resonances of single nanoparticle (shown by the dashed lines). This explains difficulties in fitting and identification of lattice resonances in Ref. [49], caused by the strong variations of interference conditions.

## 7. Conclusion

In conclusion, we theoretically studied silicon metasurfaces on the dielectric substrates with different permittivities and analyzed the role of high reflectance from the substrate. We applied a semi-analytical uncoupled-element model to describe the anti-reflecting effect, and we showed that the model is valid for all-dielectric metasurfaces on top of high-index substrates and agrees well with numerical results. We stress that the possibility of decoupling of metasurface and substrate contributions and the simplicity of the model allow understanding the main mechanism of reflection suppression from strongly scattering nanoparticles on top of the high-index substrate.

We showed that such system demonstrates broadband zero reflectance for wavelengths between the electric and magnetic dipole resonances. It originates from the destructive interference between the wave reflected from the substrate and the waves scattered by induced electric and magnetic dipole moments of the nanoparticles. We showed that this condition is satisfied for the wavelength between electric and magnetic dipole resonances when their moments are shifted in phase by π. This regime results in strong back scattering from metasurface, which allows compensating the field reflected from the substrate and achieving near-zero total reflection from the metasurfaces on top of the high-index substrate. The anti-phase Kerker condition is opposite to more common in-phase condition, which results in strong forward scattering from metasurface when electric and magnetic dipoles are in-phase.

We demonstrated that anti-reflectance effect can be also observed for nanodisk metasurface on top of the silicon substrate. For that, we proposed to decouple nanodisks from the high-index substrate by thin intermediate silica layer of 15-30 nm thickness and showed the possibility of obtaining destructive interference between metasurface- and substrate-reflected waves.

The cumulative reflection of almost independent nanoparticle scatterers provides the anti-reflectance effect and makes it tolerant with respect to inter-particle distance distortions. In particular, we designed the structure with a disordered array of nanospheres and numerically demonstrated a broadband reflection suppression: the reflectance is less than 10% at 300-650 nm wavelength range, which is better than a standard quarter-wavelength layer of silicon nitride. The main mechanism, namely the possibility of destructive interference between substrate-reflected and multipole-scattered waves, can be used to suppress reflection in either narrow or broad spectral bands. It means that size and the average density of nanoparticles influence the most, while the exact place of nanoparticles affects less.

The anti-reflectance effects addressed in this paper are supported by recent experimental results,[46-49] which among others confirms the importance of the physical model proposed in the present work. The influence of high-index substrate is crucial for designing optical metasurfaces and photovoltaic elements with nanoparticle-enhanced light trapping.

**Funding Information.** This work has been supported by the Russian Fund for Basic Research within the projects 16-52-00112 and 16-02-00684. Numerical simulations of the disordered array have been supported by the Russian Science Foundation Grant No. 16-12-10287. M.I.P. acknowledges support from Academy of Finland, Grant No. 288591.

**Acknowledgment.** We are grateful to Yuri Kivshar, Alexander Krasnok, Sergey Makarov, Arseniy Kuznetsov, and Satoshi Ishii for suggestions during manuscript preparation.

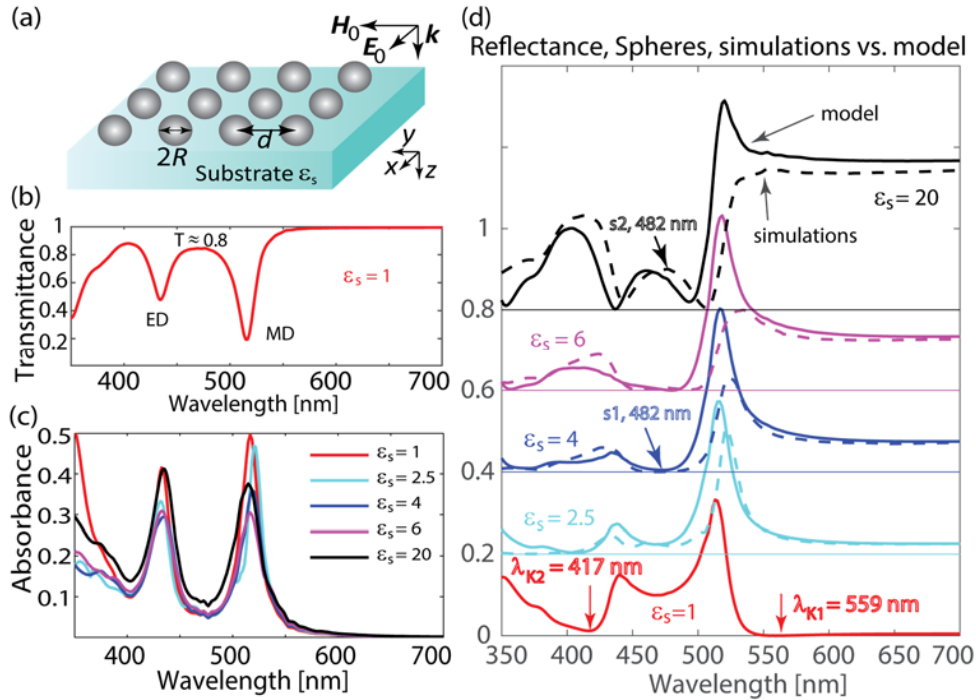

Fig. 1. (a) Schematic view of the square periodic array of nanospheres on the substrate with permittivity $\varepsilon_s$. (b) Transmittance $T = |t_{MS}^2|$ through the metasurface, that is the case $\varepsilon_s = 1$ ($d = 335$ nm and $R = 60$ nm). Between EDR and MDR, the transmittance is high (T ≈ 0.8), and results of Eqs. (2) and (3) are similar to results obtained with Eq. (S2) of Ref. [51]. However, for the wavelength of resonances, this approximation does not give good results. (c) The absorbance of the sphere array for different substrate permittivity $\varepsilon_s$ (results of the whole-structure numerical simulations, $d = 335$ nm and $R = 60$ nm). (d) The reflectance of the nanosphere metasurfaces for the same parameters as absorbance in (c). Each plot is shifted by 0.2 in respect to the previous one. Solid lines: numerical simulations. Two dips at $\lambda_{K1} = 559$ nm and $\lambda_{K2} = 417$ nm ($\varepsilon_s = 1$) correspond to the wavelengths of near-zero reflectance where the first Kerker condition is satisfied. Dashed lines: calculations according to the model where contributions of bare substrates and nanoparticle array in the air [$\varepsilon_s = 1$] calculated separately and added by Eq. (3). "s1" denotes substrate when reflectance between EDR and MDR is close to zero. Corresponding anti-phase Kerker point is shown on (d) for $\varepsilon_s = \varepsilon_{s1} = 4$ and $\lambda_A = 482$ nm. The destructive interference occurs for $2.5 < \varepsilon_s < 6$, and high reflectance is observed for the case of $\varepsilon_s = \varepsilon_{s2} = 20$, where "s2" denotes the substrate with the highest reflectance between EDR and MRD.

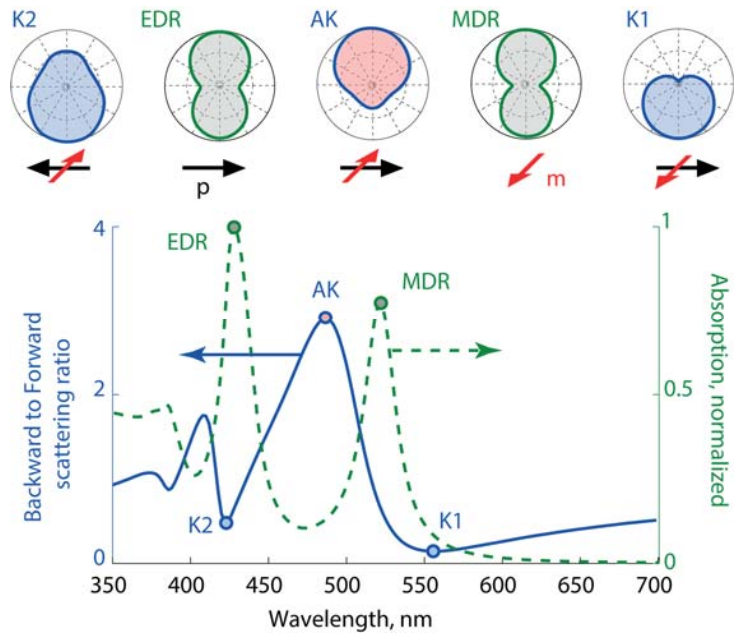

Fig. 2. The ratio of backward to forward scattered energy of individual silicon nanoparticle with $R = 60$ nm in the air compared to normalized absorption spectrum. Top inset: directivity plots for the different wavelengths of pronounced forward/backward scattering: the ratio is the highest in AK and the lowest in K1 and K2, as well as near the dipole resonances (EDR and MDR).

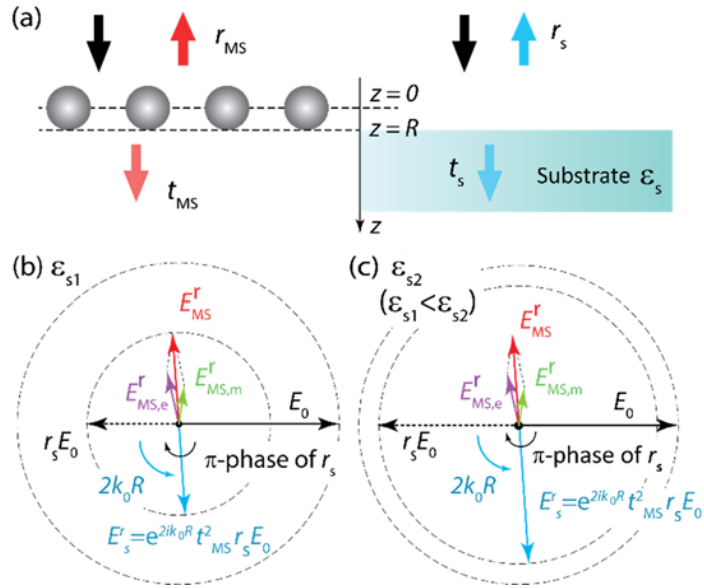

Fig. 3. (a) Decomposition of the structure into the metasurface (i.e. nanoparticle array, left) and the substrate (right). (b),(c) Artistic view of vector diagrams depicting electric field of the incident wave $E_0$ and reflected field $E_s^r$ at $z = 0$ as well as decomposition of the metasurface-reflected field $E_{MS}^r$ into contributions of electric and magnetic dipole moments $E_{MS,e}^r$ and $E_{MS,m}^r$ respectively, for two different cases: (b) $\varepsilon_s = \varepsilon_{s1} = 4$, $\lambda_A = 482$ nm (perfect match of reflected waves and near-zero reflectance) and (c) $\varepsilon_s = \varepsilon_{s2} = 20$, $\lambda_A = 482$ nm (non-zero reflectance).

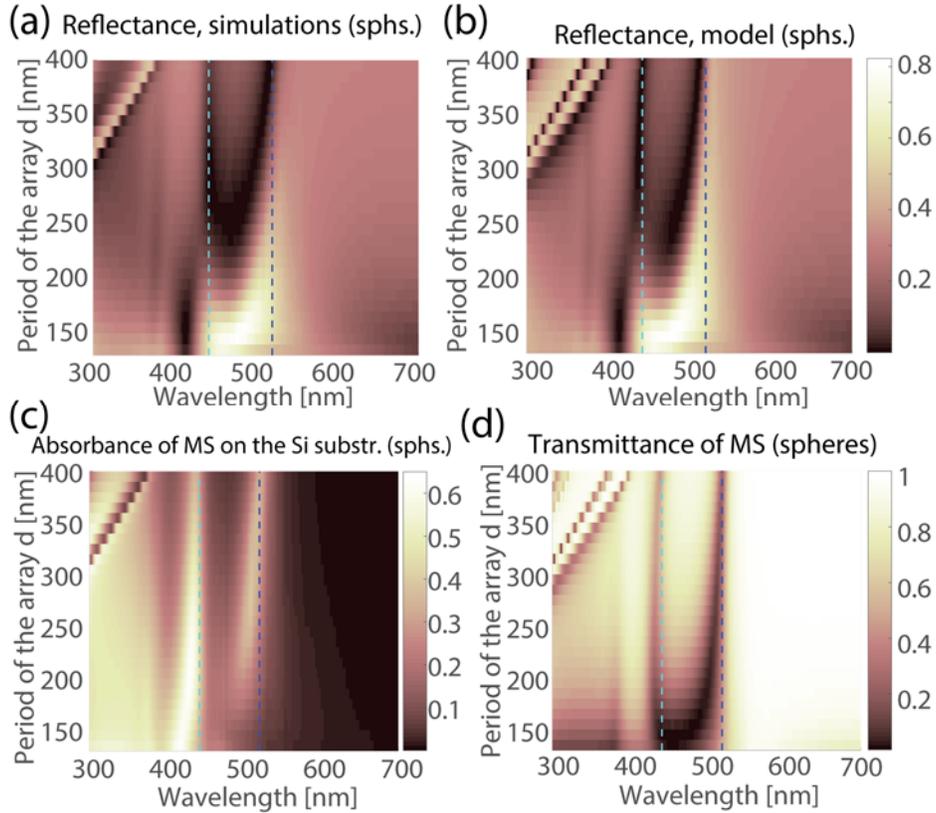

Fig. 4. Reflectance spectra of metasurfaces on Si substrate: (a) numerical simulations and (b) calculations according to the semi-analytical model (nanosphere array of $R = 60$ nm and different array periods $d = 130 - 400$ nm). Being defined as a maximum of scattering cross-section (see Fig. S7 in Ref. [51]), MDR and EDR of single nanosphere with $R = 60$ nm are shown by the dashed lines: blue and light blue respectively. Colorbar is the same for (a) and (b). (c) The absorbance of the nanosphere array on the Si substrate. (d) The transmittance of metasurface (nanosphere array without substrate).

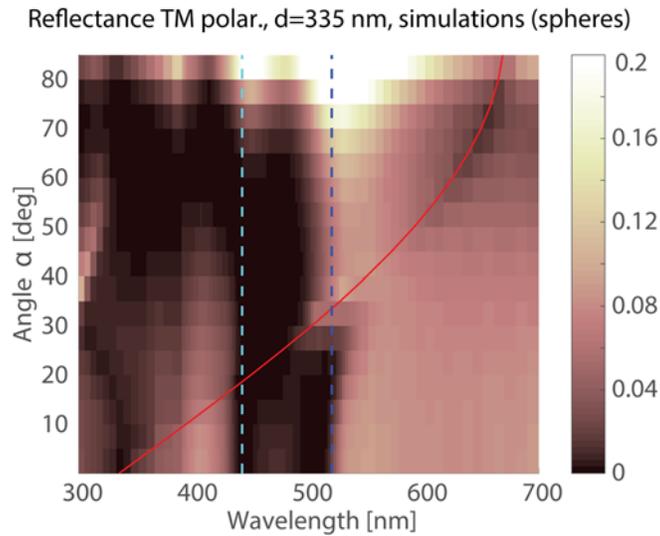

Fig. 5. Reflectance spectra of metasurface on Si substrate for different angles of incidence $\alpha$ in TM polarization (nanosphere array of $R$ = 60 nm and $d$ = 335 nm). MDR and EDR of single nanosphere are shown by the dashed lines: blue and light blue respectively. Diffraction effect defined as $\lambda_d = d\,(1 + \sin\alpha)$ is shown by the red solid line. Results for TE polarization and absorbance spectra can be found in Fig. S6 of Ref. [51].

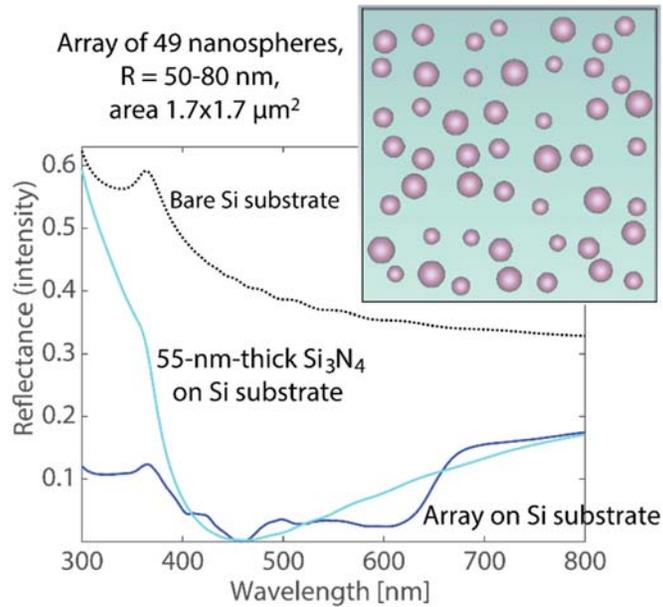

Fig. 6. Reflectance spectrum for a disordered array of silicon nanospheres on top of Si substrate (blue solid line) in comparison to reflectance from bare Si substrate (black dotted line) and from the 55-nm-thick $Si_3N_4$ layer on top of the Si substrate (light blue solid line). Inset: top view of simulation domain with 49 nanospheres, radiuses $R$ = 50-80 nm (approximately equal distribution), and the total area is 1.7 x 1.7 µm² (approximately 240 x 240 nm² per nanoparticle). The 55-nm-thick $Si_3N_4$ layer is chosen as an optimal single-layer antireflective coating that provides reflectance minimum at wavelength 470 nm and matches reflection minimum of nanoparticle array under consideration.

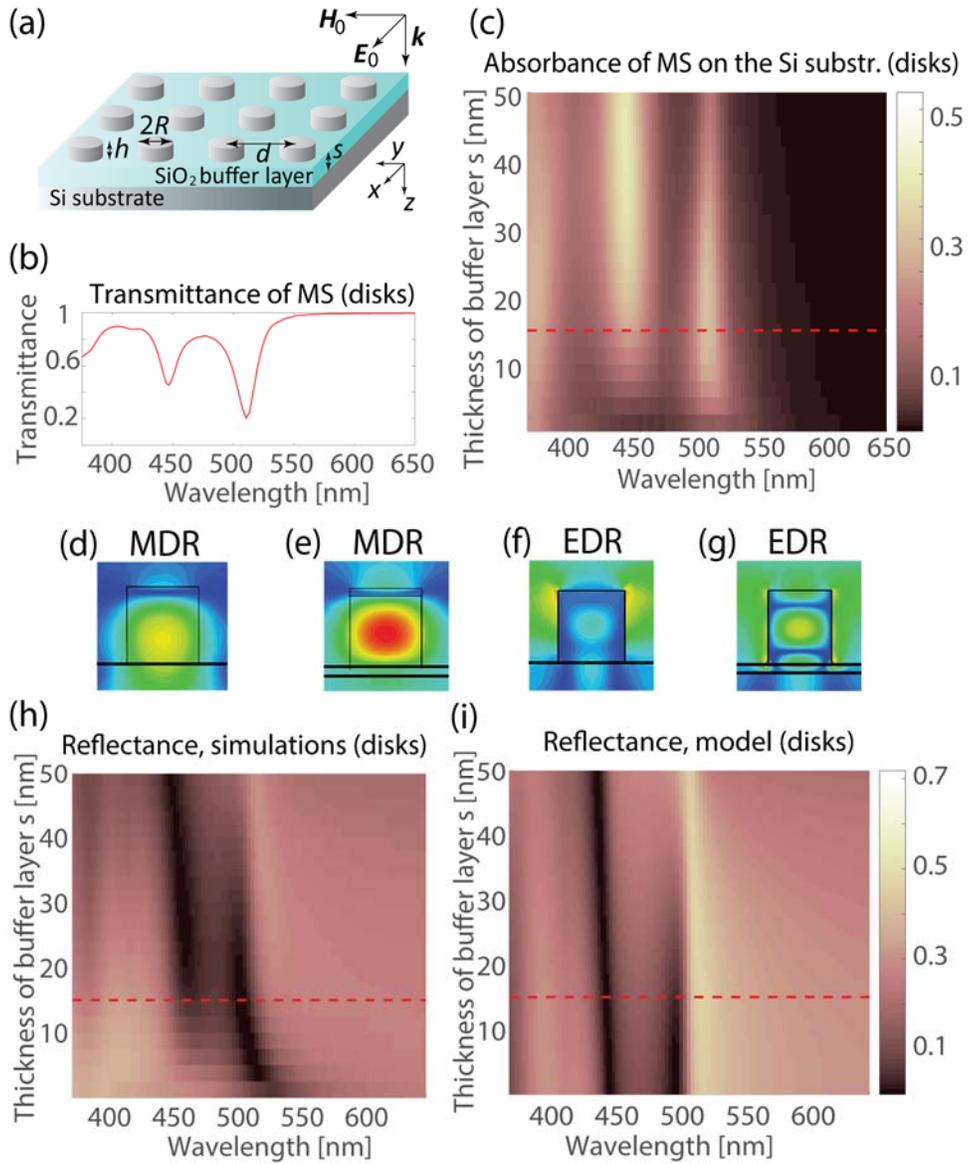

Fig. 7. (a) Schematic view of the square periodic array of nanodisks on top of the Si substrate with a thin intermediate layer of SiO$_2$ (thickness $s$). (b) The transmittance of the metasurface without substrate and buffer layer ($d$ = 350 nm). (c) The absorbance of the nanodisk array on the substrate for different thickness of the buffer layer $s$ = 0-50 nm (results of the whole-structure numerical simulations, $d$ = 350 nm). (d)-(g) Nanoparticle modes in the absence (d),(f) and presence (e),(g) of the silica buffer layer ($s_c$ = 15 nm): (d),(e) Magnetic field amplitude at the wavelength of MDR ($\lambda_{MD}$ = 513 nm) in both cases with the same colormap and (f),(g) Electric field amplitude at the wavelength of EDR ($\lambda_{ED}$ = 448 nm) in both cases with the same colormap. (h),(i) Reflectance spectra for nanodisk array on the SiO$_2$ buffer layer and Si substrate for different thickness of the buffer layer: (h) numerical simulations and (i) calculations by transfer-matrix method (colorbar is the same for both panels). Red dashed line corresponds to $s_c$ = 15 nm, so that for $s > s_c$ both resonances are presented and semi-analytical model provides results that agree well with numerical simulations.

**Supporting Information**

**Supporting-Information Figure S1**

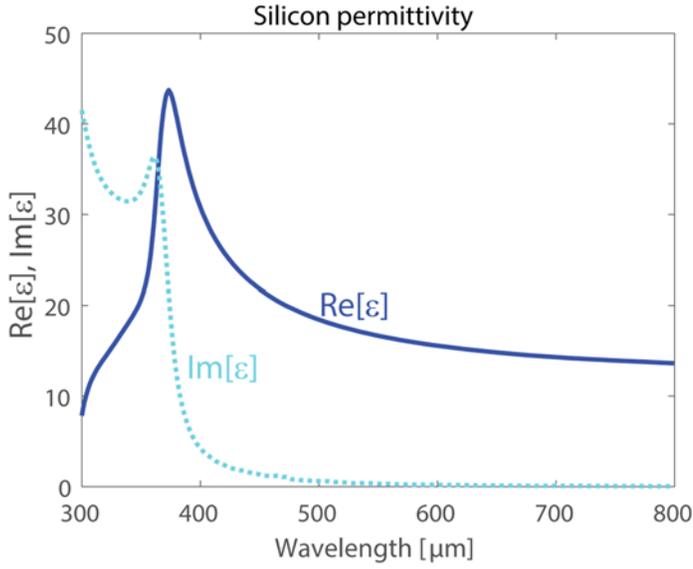

Fig. S1. Silicon permittivity used in the calculations throughout the paper. Data are from Ref. [50].

**Supporting-Information Figure S2**

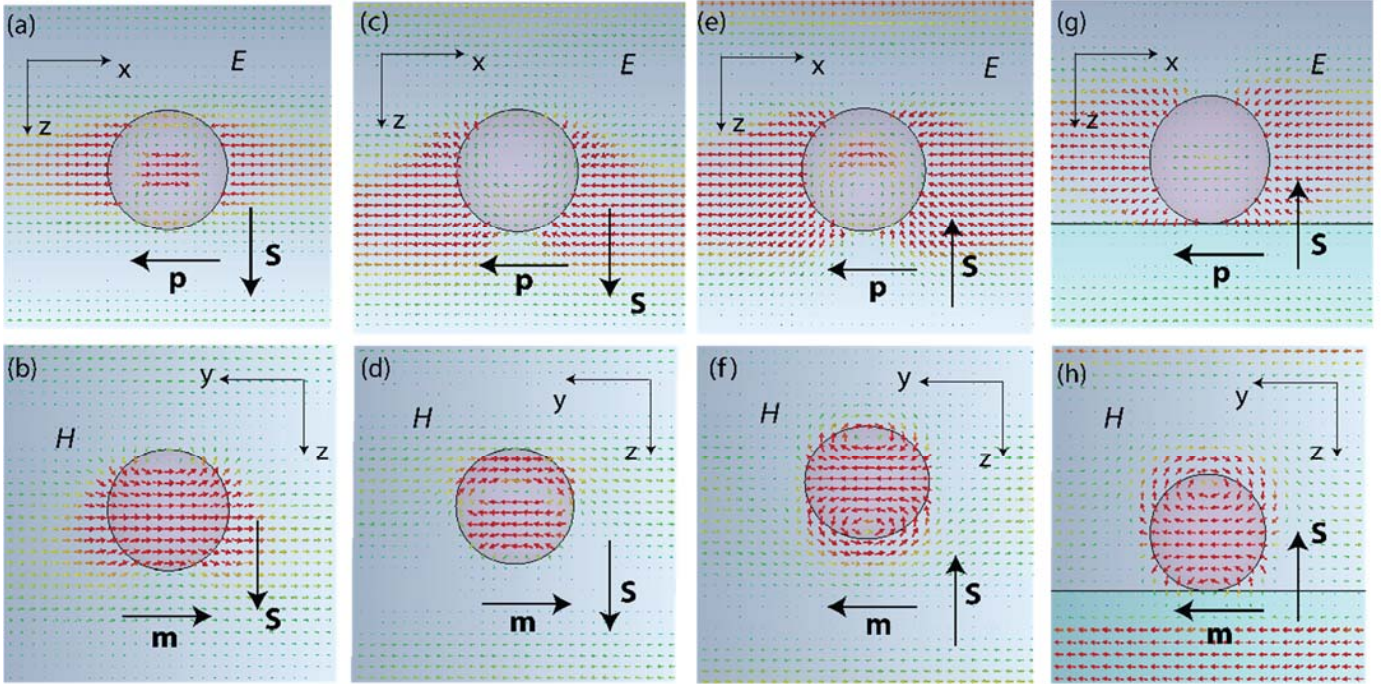

Fig. S2. Electric (top row) and magnetic (bottom row) fields distributions at $\lambda_{K1}$ = 559 nm, $\varepsilon_s$ = 1 (a),(b); $\lambda_{K2}$ = 417 nm, $\varepsilon_s$ = 1 (c),(d); $\lambda_A$ = 482 nm $\varepsilon_s$ = 1 (e),(f); and $\lambda_A$ = 482 nm, $\varepsilon_s = \varepsilon_{s1}$ (g),(h). Induced electric **p** and magnetic **m** dipole moments of nanoparticle array scatter light ( **S** $\propto [\mathbf{p} \times \mathbf{m}]$ ) either forward [**S** points down in (a)-(d)] or backward [**S** points up in (e)-(h)]. Pictures are taken at particular time moments (the same for each $E$ and $H$ pair) when induced fields are the highest. We note that **m** induced at $\lambda_{K2}$, $\varepsilon_s$ = 1 (d) is seen unclear, which agrees with the observation that Kerker effect on the blue side of the wavelength range is weaker and less pronounced than at the red side.

**Interference with the substrate**

The simplified model to take into account presence of the substrate is the first order approximation in series Eq. (2), i.e. no multiple reflection is accounted:

$$E_{\text{tot}}^{r\,(1)} = E_{\text{MS}}^{r} + E_{s}^{r\,(1)} = r_{\text{MS}} E_0 + t_{\text{MS}}^{2} r_{s} e^{2ik_0 R} E_0.\tag{S1}$$

However, one can consider even simpler case where $t_{\text{MS}}^{2} \approx 1$:

$$E_{\text{tot}}^{r\,(0)} = E_{\text{MS}}^{r} + E_{s}^{r\,(0)} = r_{\text{MS}} E_0 + r_{s} e^{2ik_0 R} E_0.\tag{S2}$$

We plot calculations with (S2) model in Fig. S3-S5 (denote as "single-reflection model") and show that it captures salient properties of the structure and give spectra similar to those obtained with the whole-structure numerical simulations. It means that destructive interference of the wave scattered by electric and magnetic dipoles and substrate-reflected wave is the main mechanism responsible for decrease of reflection between resonances.

**Supporting-Information Figure S3**

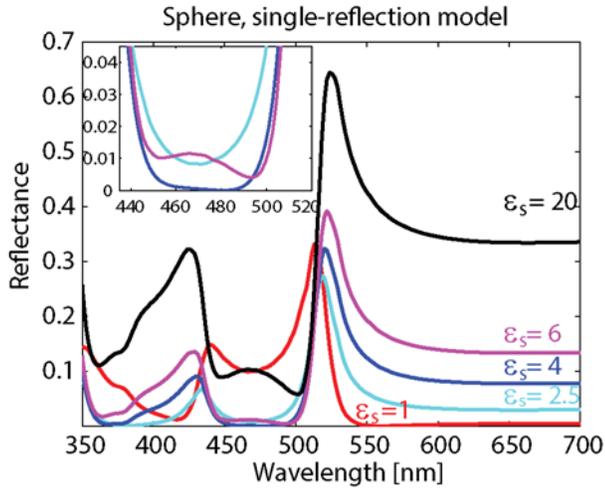

Fig. S3. Results for nanosphere array with $d = 335$ nm. Reflectance calculated within the model of independent contributions of bare substrates and nanoparticle array in air ($\varepsilon_s = 1$). In contrast to Fig. 1(d), where calculations are done using Eqs. (2) and (3), here we use single-reflection approximation (S2). Inset: enlarged view in the range 435-520 nm.

**Supporting-Information Figure S4**

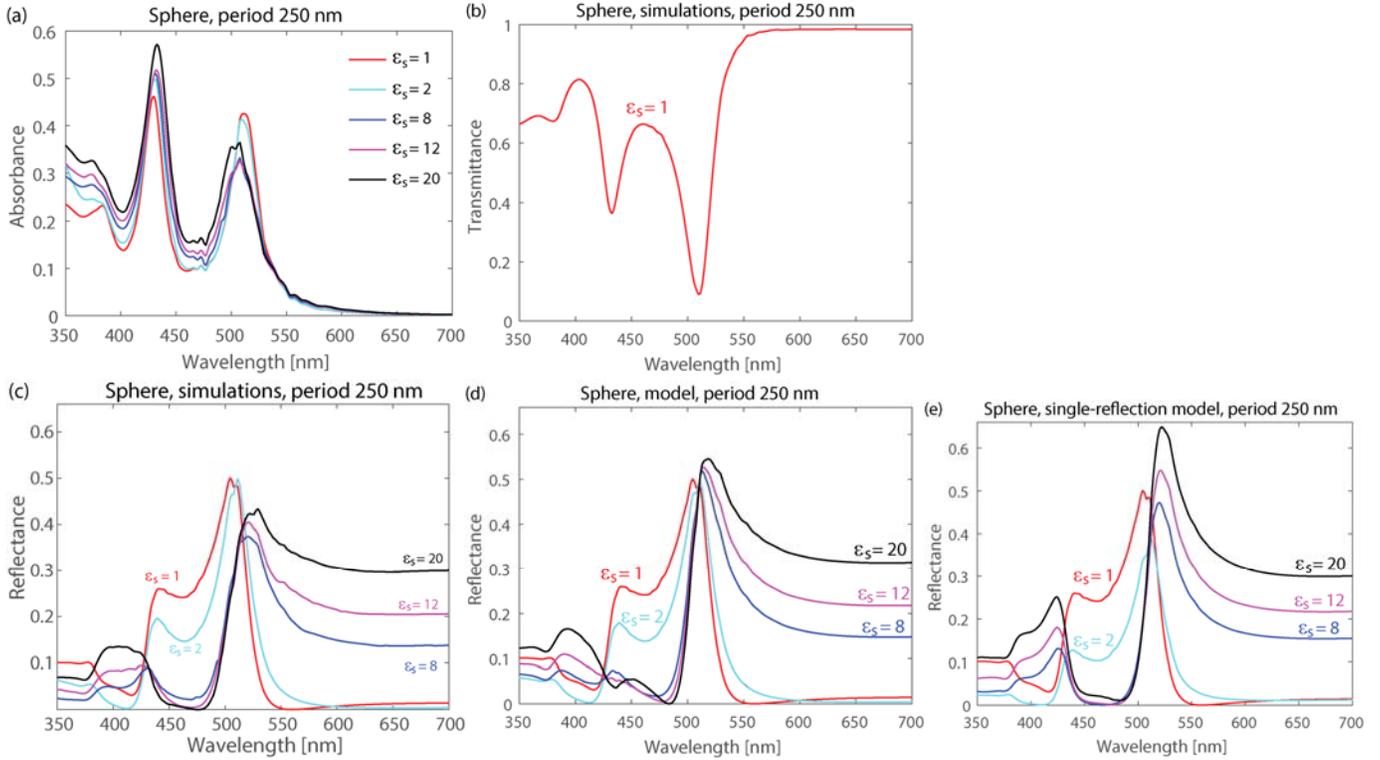

Fig. S4. Results for nanosphere array with $d$ = 250 nm. (a) Absorbance for different substrate permittivities $\varepsilon_s$ (results of the whole-structure simulations). (b) Transmittance $T = |t_{MS}^2|$ through the metasurface, that is the case $\varepsilon_s = 1$. (c) Reflectance of the same structures as (a) obtained with the whole-structure numerical simulations. (d) Calculations within the model of independent contributions of bare substrates and nanoparticle array in air ($\varepsilon_s = 1$) according to Eq. (3) (i.e. multiple reflection). (e) Calculations according to the single-reflection model Eq. (S2).

**Supporting-Information Figure S5**

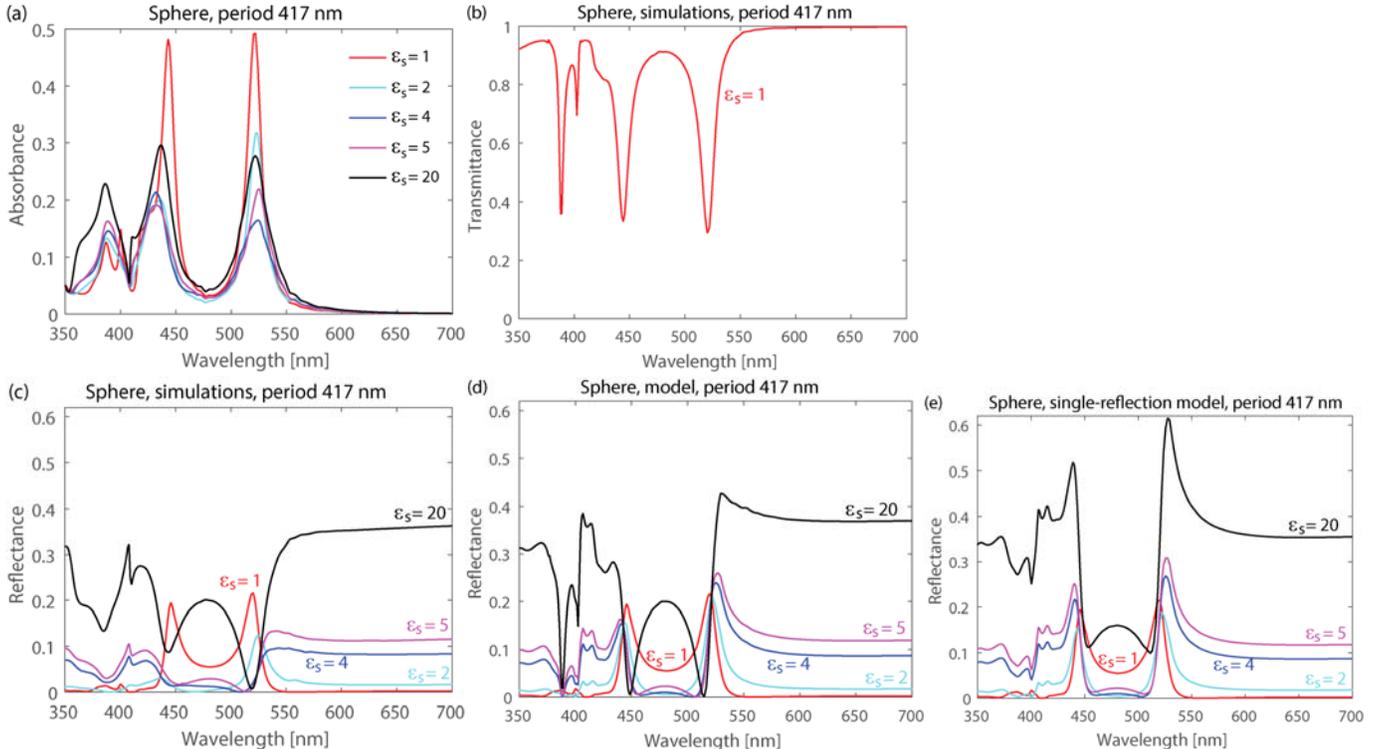

Fig. S5. The same as on Fig. S4, but for $d$ = 417 nm.

**Supporting-Information Figure S6**

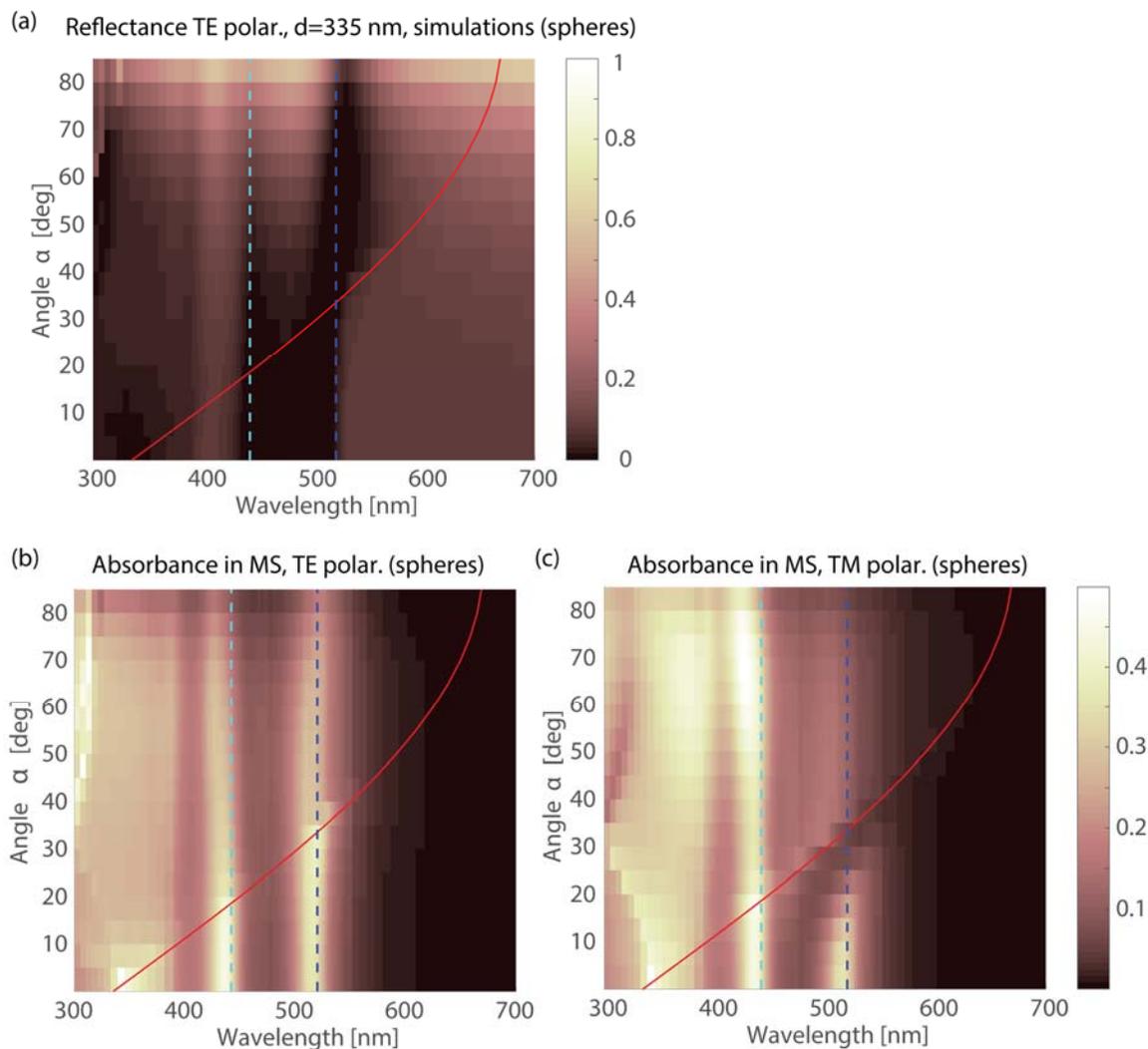

Fig. S6. (a) Reflectance spectra of nanosphere array on Si substrate for different angles of incidence $\alpha$ in TE polarization ($R = 60$ nm and $d = 335$ nm). MDR and EDR of single nanosphere are shown by the dashed lines: blue and light blue respectively. Diffraction effect defined as $\lambda_d = d(1 + \sin\alpha)$ is shown by the red solid line. (b),(c) Absorbance in metasurface for the same structure: (b) TE and (c) TM polarizations (colorbar is the same for both panels). Reflectance in TM polarization is shown in Fig. 5 of the main text.

**Supporting-Information Figure S7**

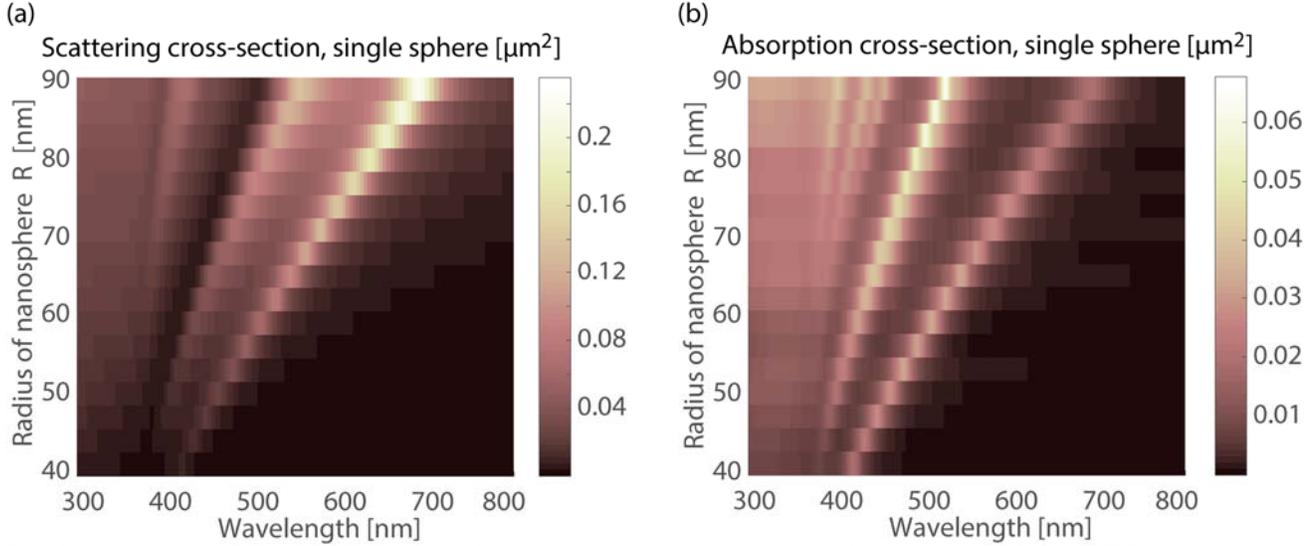

Fig. S7. (a) Scattering and (b) absorption cross-sections of single nanosphere in air. Scattering cross-section is higher for MDR than for EDR and increases with increase of nanosphere size. Absorption cross-section has the opposite tendency because of the abrupt decrease of materials loss at larger wavelength (see imaginary part of permittivity in Fig. S1).

**Transfer-matrix calculations**

Calculations of the reflectance from multilayer structure {metasurface / air gap / SiO$_2$ buffer layer / Si substrate} is performed by the following transfer-matrix approach:

$$S_{tot} = \left(\frac{T_{21}}{T_{11}}\right)^2, \text{ where } T_{21} \text{ and } T_{11} \text{ are matrix elements in } T_m = \begin{bmatrix} T_{11} & T_{12} \\ T_{21} & T_{22} \end{bmatrix},$$

and $T_m = T_{MS} T_{p1} T_{i1} T_{p2} T_{i2}$ with $T_{p1} = \begin{bmatrix} e^{ik_0 h/2} & 0 \\ 0 & e^{-ik_0 h/2} \end{bmatrix}$, $T_{p2} = \begin{bmatrix} e^{ik_b s} & 0 \\ 0 & e^{-ik_b s} \end{bmatrix}$, $T_{MS} = \frac{1}{t_{MS}} \begin{bmatrix} 1 & -r_{MS} \\ r_{MS} & t_{MS}^2 - r_{MS}^2 \end{bmatrix}$, $T_{i1} = \frac{1}{t_{i1}} \begin{bmatrix} 1 & r_{i1} \\ r_{i1} & 1 \end{bmatrix}$,

$T_{i2} = \frac{1}{t_{i2}} \begin{bmatrix} 1 & r_{i2} \\ r_{i2} & 1 \end{bmatrix}$, $r_{i1} = \frac{1-\sqrt{\varepsilon_{SiO2}}}{1+\sqrt{\varepsilon_{SiO2}}}$, $t_{i1} = \frac{2\left(\sqrt{\varepsilon_{SiO2}}\right)^{1/2}}{1+\sqrt{\varepsilon_{SiO2}}}$, $r_{i2} = \frac{\sqrt{\varepsilon_{SiO2}} - \sqrt{\varepsilon_{Si}}}{\sqrt{\varepsilon_{SiO2}} + \sqrt{\varepsilon_{Si}}}$, $t_{i2} = \frac{2\left(\sqrt{\varepsilon_{SiO2}}\sqrt{\varepsilon_{Si}}\right)^{1/2}}{\sqrt{\varepsilon_{SiO2}} + \sqrt{\varepsilon_{Si}}}$, $k_0 = 2\pi/\lambda$,

$k_b = k_0 \sqrt{\varepsilon_{SiO2}}$.

For the case $s = 0$ (no SiO$_2$ buffer layer), analytical expression for $S_{tot}$ is the same as Eq. (3) gives.

**Nanopillar array on the high-index substrate without buffer layer**

We performed numerical simulations and comparison for the nanodisk array directly on top of high-index substrate with varied permittivity [Fig. S8(a)]. Because of the strong coupling between nanoparticle and the substrate, the amplitude of waves diffracted in the substrate becomes high in the case of lossless substrate. To suppress wave propagation inside the substrate and eliminate diffraction effects, we made simulations with small imaginary part of the complex permittivity $\varepsilon_s = \varepsilon_s' + i\varepsilon_s''$, in particular $\varepsilon_s'' = \varepsilon_s'/20$.

Under increase of $\varepsilon_s'$, both ED and MD peaks are vanished [Fig. S8(b)]. EDR disappears almost completely, and for both peaks, it happens because of the increasing coupling between the nanodisk and the substrate and consequent mode leakage. Similar to the array of nanospheres, for nanopillar array, one can observe the presence of antireflection between EDR and MDR [Fig. S8(c)]. However, in contrast to spheres, the lowest point $\lambda_A$ experiences a redshift, the reflectance is minimal for $\varepsilon_s' = 8$, and further increase of $\varepsilon_s'$ causes an increase of reflectance.

Comparing the simulation results with the calculations according to the model of separate contributions [Fig. S8(d)], we see the tendency which is similar to the sphere array: antireflection point takes place between the EDR and MDR, the lowest value are observed for $\varepsilon_s' = 8$, followed by an increase of the reflectance. However, the model does not predict the change of wavelength $\lambda_A$. Indeed, the interaction of the nanodisk modes with the substrate is strong [see suppression of the disk's EDR and MDR in Fig. S8(b)], and our approach based on separation of the metasurface and substrate contributions does not work well.

**Supporting-Information Figure S8**

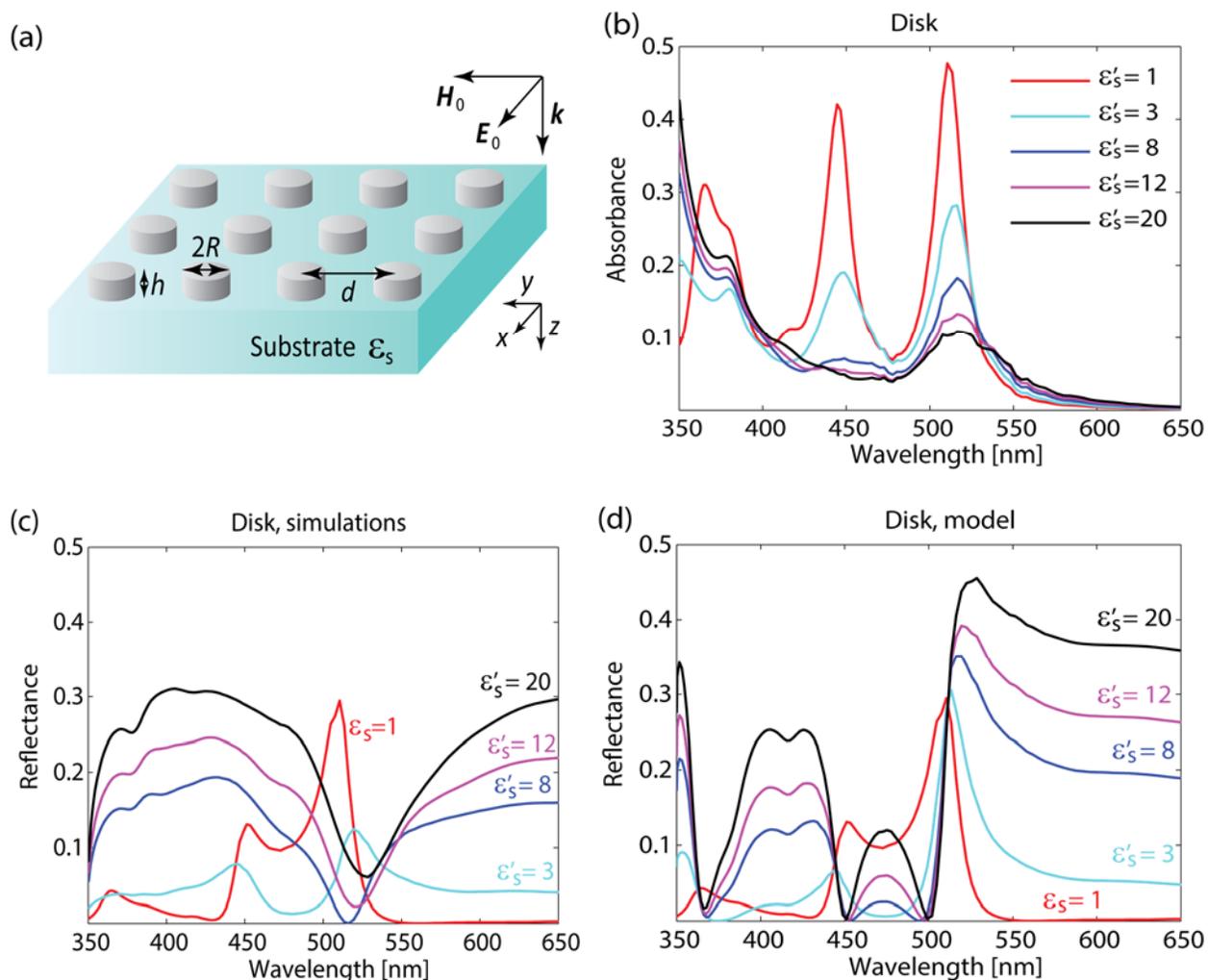

Fig. S8. (a) Schematic view of the square periodic array of nanodisks on the substrate with permittivity $\varepsilon_s$. (b) Absorbance of the nanodisk array for different substrate permittivity $\varepsilon_s$ (results of the whole-structure numerical simulations, $d = 350$ nm). (c) Reflectance of the disk metasurfaces for the same parameters as absorbance (numerical simulations). (d) Calculations according to the model where contributions of bare substrates and nanodisk array in air [$\varepsilon_s = 1$] calculated separately and added by Eq. (3).